\documentclass[twocolumn]{aastex631}
 \usepackage{ifpdf}
 \usepackage{graphicx, subfigure}
 \usepackage{natbib}
 \usepackage{times}
 \usepackage{amsmath}
 \usepackage{textcomp}
 \usepackage[colorinlistoftodos,prependcaption,textsize=tiny]{todonotes}
 \usepackage{xcolor}
 \usepackage{gensymb}

\newcommand{\bbe}{\begin{eqnarray}}
\newcommand{\ee}{\end{eqnarray}}

\begin{document}
\nolinenumbers

\title{XMM--NuSTAR Observation and Multiwavelength SED Modeling of Blazar 4FGL J1520.8--0348}

\author{Garima Rajguru}
\affil{Department of Physics and Astronomy, Clemson University, Kinard Lab of Physics, Clemson, SC 29634-0978, USA}
\email{grajgur@clemson.edu}

\author{L. Marcotulli}
\altaffiliation{NHFP Einstein Fellow}
\affil{Yale Center for Astronomy \& Astrophysics, 52 Hillhouse Avenue, New Haven, CT 06511, USA}
\affil{Department of Physics, Yale University, P.O. Box 208120, New Haven, CT 06520, USA}
\email{lea.marcotulli@yale.edu}

\author{M. Ajello}
\affil{Department of Physics and Astronomy, Clemson University, Kinard Lab of Physics, Clemson, SC 29634-0978, USA}

\author{A. Tramacere}
\affil{Department of Astronomy, University of Geneva, Ch. d’Ecogia 16, Versoix, 1290, Switzerland}




\begin{abstract}
\nolinenumbers
Active galactic nuclei (AGNs) can power relativistic jets, which are called blazars when pointed close to our line of sight. Depending on the presence or absence of emission lines in their optical spectra, blazars are categorized into flat spectrum radio quasars (FSRQs) or BL Lacertae (BL Lacs) objects. According to the `blazar sequence', as synchrotron peak frequency ($\nu^{sy}_{pk}$) shifts to higher energies, the synchrotron peak luminosity  decreases. This means that BL Lacs as luminous as FSRQs, and with synchrotron peak frequencies $\nu^{sy}_{pk}>10^{15}$ Hz, should not exist. Detected as a high-synchrotron peak (HSP; $\nu^{sy}_{pk}>10^{15}$ Hz) BL Lac, 4FGL J1520.8-0348 shows high gamma-ray luminosity ($L_{\gamma}>10^{46}\,\rm erg~s^{-1}$), being at a high redshift of $z=$1.46. Since it is an outlier in the `blazar sequence', the process of its jet acceleration and power may be different from bona fide BL Lacs. In this work, we constrain its spectral energy distribution (SED) by modeling the multi-wavelength data from infrared to $\gamma$-ray regime. Simultaneous X-ray data was obtained from \textit{X-ray Multi-Mirror Mission} and \textit{Nuclear Spectroscopic Telescope Array} to constrain the synchrotron emission and underlying electron distribution. On undertaking the SED modeling of the source, including the effect of extragalactic background light, we conclude that the source is more likely to be a `blue FSRQ' or `masquerading BL Lac' where the BL Lac is actually a FSRQ in disguise.
\end{abstract}

\section{Introduction} 
\label{sec:intro}

Blazars are an extreme class of active galactic nuclei (AGNs), whose jets are closely aligned towards our line of sight \citep{UrryPadovani1995}. Blazar jets are broadband (radio to gamma) emitters and their spectral energy distribution (SED) shows a double-humped structure \citep[e.g.,][]{Padovani2017}. The lower energy hump (from radio to UV or X-rays) is due to synchrotron emission from relativistic electrons present in the jet. The high energy peak (keV-GeV) is due to inverse Compton (IC) scattering of the same jet photons to higher energies by the energetic jet electrons, a process called the synchrotron self-Compton \citep[SSC,][]{Maraschi1992, Bloom1996, Ghisellini1985}. In case the jet electrons upscatter external photons from the broad line region (BLR), the dusty torus (DT) and/or the accretion disk, the process is called external Compton \citep[EC,][]{Sikora1994, Dermer1994, EC_DT, Ghisellini2009}. Therefore, collecting and modeling the broadband SED of a blazar is crucial to understand the underlying physics of the source.

Blazars can be classified into two subclasses based on the presence (or absence) of broad (equivalent width, $EW>5$\AA) emission lines in the optical spectra \citep{UrryPadovani1995}. It is understood that the BLR clouds are photo-ionized by the disk radiation. Recombination in ionized clouds produce the broad emission lines. If the optical spectra show prominent emission lines, the blazars are termed as flat spectrum radio quasars (FSRQs). If the spectrum is devoid of broad emission lines (EW $<$ 5\AA), they are called BL Lacertae (BL Lac) objects.
Depending on the position of the synchrotron peak frequency ($\nu^{sy}_{pk}$), i.e. the lower energy peak, blazars are classified in 3 categories. If $\nu^{sy}_{pk}<10^{14}$ Hz it is a low-synchrotron peak blazar (LSP) whereas if $\nu^{sy}_{pk}>10^{15}$ Hz then it is a high-synchrotron peak blazar (HSP). Intermediate-synchrotron peak blazars (ISP) have $10^{14} <\nu^{sy}_{pk}<10^{15}$ Hz \citep{Abdo2010}. The `blazar sequence' discussed in \citet{BlazaeSeqFossati1998} suggests that the peak frequency of both the synchrotron and inverse Compton peaks shifts to lower energies with increasing peak luminosity. Usually FSRQs are more luminous than BL Lacs, and constitute the majority of the LSP population. HSP BL Lacs have even been found with $\nu^{sy}_{pk}>10^{17}$ Hz \citep{Ackermann2015}, accelerating particles to TeV range \citep{Costamante2001, Tavecchio2011}.


Utilizing various methods \citep{Rau2012, Shaw2013, Ajello2014, Shaw2014, kaur2017}, the redshifts of $\sim$200 brightest BL Lacs detected by the $Fermi$ Large Area Telescope \citep[LAT][]{4fgl} have been obtained. Sixteen BL Lacs were found to be at high redshifts $z>1$ \citep{Rau2012,kaur2017}, which indicates that their luminosity is comparable to FSRQs (e.g. bolometric luminosity, $L_{\rm bol}>10^{46}\,\rm erg~s^{-1}$ \citep{Ghisellini2017_blazarseq}). However, they also possess hard $\gamma$-ray spectra (photon index, $\Gamma_{\gamma}<2$), they are detected up to $E>60\,\rm GeV$, and have high-syncrotron peak frequencies ($\nu_{synch}>10^{15}\,\rm Hz$), akin to HSP BL Lacs. Such sources which exhibit a dichotomy in their observed properties have been referred to as `blue FSRQs' or `masquerading BL Lacs'\citep{blueFSRQ,Padovani2012,Giommi2013,Rajagopal2020}. The physical interpretation is that they could be powerful FSRQs whose jets swamp any emission lines in the optical spectra. \citet{TXS0506+056} discussed various parameters implying similar dichotomy in the source TXS 0506+056, which has also been known to emit high-energy neutrinos \citep{Icecube18}. 

4FGL J1520.8-0348 (hereafter J1520) is one such source, which is the object of interest in our work. Using the photometric redshift technique \citet{kaur2017} found the redshift of J1520 to be at $z=1.46^{+0.12}_{-0.11}$. In order to probe the nature of this source, we carried out the multiwavelength SED modeling. Since the synchrotron peaks in X-ray energies for HSP BL Lacs, to constrain the synchrotron peak more precisely, simultaneous data from X-ray Multi-Mirror Mission \citep[\it XMM-Newton;][]{xmm} and Nuclear Spectroscopic Telescope Array \citep[\it NuSTAR;][]{nustar} was acquired (NuSTAR joint XMM-Newton General Observer cycle-8 program, Proposal number 8135) and analysed. Furthermore, data from the $Fermi$-LAT \citep{4fgl}, Gamma-Ray Optical/Near-infrared Detector (GROND) \citep{GROND} and archival data was added to construct the multiwavelength SED spanning IR to $\gamma$-rays. SED modeling was undertaken to constrain jet parameters which would shed some light on the underlying physical processes of the source.

This paper is organized in the following manner. In section \ref{sec:target} we discuss our source selection. In section \ref{sec:obs} we elaborate on the gathered data and its analysis. The X-ray spectral analysis has been described in section \ref{sec:xray} and in section \ref{sec:sed} the SED modeling of multiwavelenth data has been elucidated. The main discussion and conclusion has been summarized in section \ref{sec:conclusion}. Throughout this paper, we use a flat $\Lambda$CDM cosmological model with $H_0=67.8 \rm ~km~s^{-1}~Mpc^{-1}$, $\Omega_m = 0.30$ and $\Omega_{\Lambda}=0.69$.

\begin{table*}[htb!]
\large
\caption{Analysis results and model parameters obtained from optical, UV and X-ray analysis}
\resizebox{\textwidth}{!}{
\begingroup
\setlength{\tabcolsep}{10pt}
\renewcommand{\arraystretch}{1.5}
\begin{tabular}{ c c c c c c }
\tableline
\tableline
\hline
\multicolumn{6}{c}{\bf GROND AB Magnitudes (Optical)}\\
\tableline
\tableline
$g\arcmin$ & $r\arcmin$ & $i\arcmin$& $z\arcmin$ & &\\
\hline
$17.61\pm0.03$& $17.25\pm0.04$ & $17.13\pm0.04$ & $16.94\pm0.04$ & &\\
\tableline
\tableline
\multicolumn{6}{c}{\textbf{GROND AB  Magnitudes (IR)}} \\
\tableline
\tableline
$J$ & $H$ & $K_s$\\
\hline
$16.65 \pm 0.05$ & $16.31 \pm 0.08$ & $16.05 \pm 0.10$ \\
\tableline
\tableline
\multicolumn{6}{c}{\textbf{\textit{Fermi}-LAT}}\\
\tableline
\tableline
$\Gamma_{\gamma}^{a}$ & $\beta_\gamma^a$ & Flux$^{b}$ ($10^{-11}\,\rm erg~cm^{-2}~s^{-1}$) & Counterpart & Radio Luminosity ($10^{42}\,\rm erg~s^{-1}$) & \\
\hline
$1.55\pm0.07$ & $0.15\pm0.03$ & $1.14\pm0.08$ & NVSS J152048$-$034850 & $8.98\pm0.28$ &\\
\tableline
\tableline
\multicolumn{6}{c}{\textbf{\textit{XMM + NuSTAR$^{e}$}}}\\
\tableline
\tableline
$\alpha_{\rm X}^{c}$ & $\beta$ & $E_{\rm pivot}\,\rm (keV)$& Flux$^{d}$ ($10^{-12}\,\rm erg~cm^{-2}~s^{-1}$) & reduced $\chi^2_{\nu}$ (D.O.F.) &\\
\hline
$2.73\pm0.07$ & $0.54\pm0.24$  & $1.0$ & $1.62\pm0.02$ &  $1.017~(109)$ & \\
\tableline
\tableline
\tableline
\multicolumn{6}{l}{\begin{minipage}{\textwidth} 
\tablenotetext{a}{Log-parabolic $\gamma$-ray index and curvature $\beta_\gamma$ from 4FGL-DR4.}
\tablenotetext{b}{$\gamma$-ray flux between $1-100\,\rm GeV$ from 4FGL-DR4.}
\tablenotetext{c}{X-ray log-parabola index obtained from XSPEC analysis.}
\tablenotetext{d}{Integrated X-ray flux from 0.2 to 20 keV obtained from XSPEC analysis. This is the unabsorbed value of flux obtained after correcting for the Galactic absorption.}
\tablenotetext{e}{The errors of all parameters obtained from XSPEC fitting are at the 90\% confidence level.}
\end{minipage}%
}
\end{tabular}
\endgroup
}
\label{Tab:d1}
\end{table*}


\section{TARGET SELECTION}\label{sec:target}
4FGL J1520.8-0348, also known as NVSS J152048-034850, is a high-$z$ HSP BL Lac, found to be at $z=1.46^{+0.12}_{-0.11}$ by \citealp{kaur2017}. This source is bright in X-rays, and was detected for the first time by the {\it ROSAT} All Sky Survey \footnote{\url{https://tools.ssdc.asi.it/SED/}} \citep[RASS][]{RASS} at a flux level of $F_{0.1-2.4\,\rm keV}=(5.7\pm1.9)\times 10^{-13}\,\rm erg~cm^{-2}~s^{-1}$. Moreover, it was observed by \textit{Swift/XRT} \citep{Swift_xrt} in 2009 (obsid: 00039185001, 8.28 ks) \footnote{\url{https://www.swift.ac.uk/swift_live/index.php}}  with a measured flux of $F_{0.3-10.0\,\rm keV}=6.5_{-0.6}^{+0.9}\times10^{-13}\,\rm erg~cm^{-2}~s^{-1}$. On comparing these flux levels, and those that we obtained from our X-ray data (see Sec. \ref{sec:sed}), we see that the source has not shown variability in X-rays. The 1FGL catalog \citep{1fgl} and subsequent {\it Fermi}-LAT catalogs \citep{2fgl, 3fgl, 4fgl} detected J1520 in $\gamma$-ray regime, as a non-variable source with a Variability Index of 23.63 \citep[$<$27.69 threshold, ][]{4fgl_dr4, Fermi_lc}. It possesses a $\Gamma_{\gamma}\sim1.78(\pm0.03)$ \citep[4FGL-DR4,][]{4fgl_dr4}, exhibiting a hard $\gamma$-ray spectrum, with $\gamma$-ray luminosity $>10^{47}\,\rm erg~s^{-1}$ (band: 0.1-100 GeV). It is an object of interest as a possible `masquerading BL Lac' since it displays properties of both FSRQs and BL Lacs. Its BL Lac nature is evinced by the absence of spectral lines in the optical regime and Compton Dominance $<1$ \citep{shaw2013b,paiano2017}. Compton dominance is the ratio of inverse Compton luminosity to synchrotron luminosity. It is $\le 1$ for BL Lacs and $>1$ for FSRQs. On the contrary, a high radio power\footnote{\url{https://ned.ipac.caltech.edu/}} ($P_{1.4GHz}\sim10^{26}$ W Hz$^{-1}$), possibly high Eddington ratio \citep[L$_{disk}$/L$_{EDD} \sim$ 0.03 ][]{kaur2017} and high synchrotron peak luminosity ($L_{sy}^{pk}>10^{46}\,\rm erg~s^{-1}$) indicates that the source may also be a FSRQ.

\section{OBSERVATIONS AND DATA ANALYSIS}\label{sec:obs}

\subsection{{\it Fermi}-LAT}
The $\gamma$-ray spectrum of J1520 was obtained from the fourth $Fermi$-LAT source catalog - Data Release 4 \citep[4FGL-DR4;][]{4fgl, 4fgl_dr4}. The energy flux (100 MeV - 100 GeV) obtained by spectral fitting ($F_\gamma=1.14(\pm0.08) \times 10^{-11} \, \rm erg~cm^{-2}~s^{-1}$) and log-parabolic index of the source (1.55$\pm$0.07), as determined by 4FGL-DR4 catalog, is presented in Table \ref{Tab:d1}.
\subsection{{\it NuSTAR}}

{\it NuSTAR} observation of J1520 was made on January 21, 2023 for an exposure time of 54.6 ks. {\it NuSTAR} Data Analysis Software ({\it NuSTARDAS}), consolidated in the HEASoft v.6.31.1\footnote{\url{https://heasarc.gsfc.nasa.gov/docs/software/lheasoft/}} software package, was utilised to analyse data from Focal Plane Modules A (FPMA) and B (FPMB) instruments aboard {\it NuSTAR}. Using the response files from {\tt\string CALDB} database (v.20230504), the {\tt\string nupipeline} task was employed for data reduction. A circular region of radius 15$\arcsec$ centered on the source coordinates was selected for source extraction and a circular region of radius 30$\arcsec$ was selected for the background region. The background region was ensured not to be contaminated by the source but selected in the same frame as the source. Resulting spectra and matrix files were generated using the {\tt\string nuproducts} task. Rebinning was done on the spectral files to contain at least 10 counts per bin.

\subsection{{\it XMM-Newton}}

{\it XMM-Newton} observed J1520 on January 21, 2023 (simultaneously with {\it NuSTAR}) for an exposure time of 14.4 ks. {\it XMM-Newton} consists of three European Photon Imaging Cameras (EPIC): MOS1, MOS2 (Metal Oxide Semiconductor) and pn CCD, with an energy sensitivity range of 0.2 to 12 keV. {\it XMM-Newton} Science Analysis Software (SAS) v20.0.0\footnote{\url{https://www.cosmos.esa.int/web/xmm-newton/what-is-sas}} was used to process the data and tasks {\tt\string emproc} and {\tt\string epproc} were implemented to generate the event files for MOS and pn respectively. In this case, a circular region of radius 10$\arcsec$ centered on the source coordinates, was selected for source extraction, a circular region of radius 15$\arcsec$ was selected for the background region and the task {\tt\string evselect} was used to generate the corresponding spectra. Rebinning was done on the spectral files to contain at least 25 counts per bin.

\subsection{GROND \& Archival Data}
Optical-IR data for J1520 was obtained in 7 different bands ($g' , r' , i' , z', J, H, K_s$) by \citealp{kaur2017}, using Gamma-Ray Optical/Near-infrared Detector (GROND) mounted at the MPG 2.2m telescope at European Southern Observatory (ESO) La Silla, Chile \citep{GROND}. Table \ref{Tab:d1} reports the AB magnitudes obtained from \citealp{kaur2017}.
Archival data in the infrared regime was obtained from the SED Builder Tool\footnote{\url{https://tools.ssdc.asi.it/SED/}} of Space Science Data Center (SSDC).



\begin{figure}[htb!]
    \centering
	\makebox{\includegraphics[width=\columnwidth]{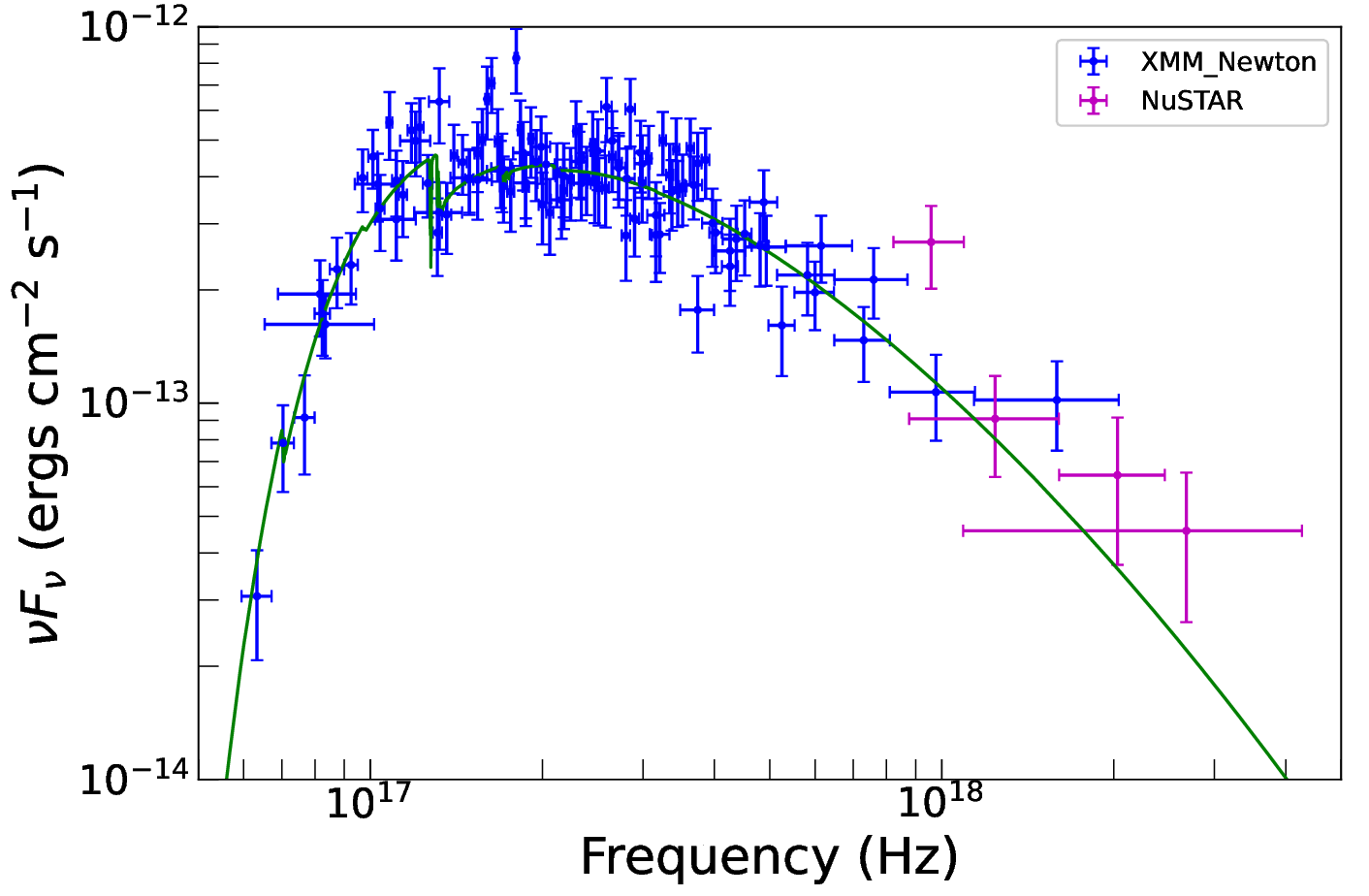}}	
		\caption{\label{fig:xray} X-ray spectral analysis using Xpsec to fit a log-parabola model to {\it XMM-Newton} and {\it NuSTAR} data, corrected for Galactic absorption. The green solid line is the fitted model.
		}
\end{figure}

\section{X-ray Spectral Analysis}
\label{sec:xray}

The {\tt\string XSPEC} tool, integrated within the HEASoft package was employed to analyze the spectral data from {\it XMM-Newton} and {\it NuSTAR}. Galactic column density (N$_{H}$) along the line of sight was obtained from the High Energy Astrophysics Science Archive Research Center (HEASARC) Website\footnote{\url{https://heasarc.gsfc.nasa.gov/cgi-bin/Tools/w3nh/w3nh.pl}}, and was found to be $N_{H}=7.28 \times 10^{20}$ cm$^{-2}$ \citep{Kalberla2005}. A log parabola model (\texttt{logpar} in {\tt\string XSPEC}) multiplied with a constant factor to take into account the cross-calibration between the {\it XMM-Newton} and {\it NuSTAR} data (\texttt{const}), corrected for galactic absorption (\texttt{tbabs}), was fit using $\chi^2$ statistics onto the {\it XMM-Newton} and {\it NuSTAR} source spectra freezing $N_{\rm H}$ to $7.28 \times 10^{20}$ cm$^{-2}$, and keeping all other parameters free. We noted that there is a constant factor of 1.2 between the {\it XMM-Newton} data and {\it NuSTAR} data, which is consistent with cross-calibration between the two instruments \citep{TelescopeCrossCalibration}. The result of this analysis is shown in Table \ref{Tab:d1}.

Figure \ref{fig:xray} shows the result of the fit. The combined X-ray spectra was also fitted with a simple power-law (\texttt{powerlaw}) and broken power-law model (\texttt{bknpo}), for comparison. The reduced $\chi^{2}$ values obtained were 1.14 and 1.03 respectively, both of which are greater than reduced $\chi^{2}=$ 1.017 for log-parabolic model. On operating \texttt{ftest} on log-parabolic model and simple power law, we obtain the F statistic value $=$ 15.431 and Probability $\sim$ 10$^{-4}$. Also, the confidence range of errors associated with the parameters of each model was most consistent in case of log-parabolic model, making it the best-fit model.



\section{SED Modeling}
\label{sec:sed}
To understand the physical properties of the source, we modeled the multiwavelength data using a one-zone leptonic model. The emitting region is assumed to be spherical, located at a certain distance ($\rm R_{diss}$) from the central supermassive black hole (described by its mass, $M_{\rm BH}$). The electron population underlying the emission has an intrinsically curved spectrum (resulting from energy dissipation processes that affected the region during its path along the jet) that can be described either by a log-parabolic shape of the form:

\begin{equation}
    n(\gamma) = n_0
   \left(\frac{\gamma}{\gamma_{\rm break}}\right)^{-[s+r\log(\gamma/\gamma_{\rm break})]} \quad \gamma_{\rm min}\leq \gamma \leq\gamma_{\rm max}
\label{eq:ele_dis_llp}
\end{equation}

In the above, $\gamma$ are the electron energies in the range $\gamma_{min}$ to $\gamma_{max}$ (where $\gamma=1/\sqrt{1-\beta^2}$, and $\beta=v/c$); $s$ is the power-law slope of the electron distribution; $\gamma_{\rm break}$ is the turnover energy; and $r$
is the curvature of the distribution. 

\begin{figure}[ht!]
    \centering
	\makebox{\includegraphics[width=\columnwidth]{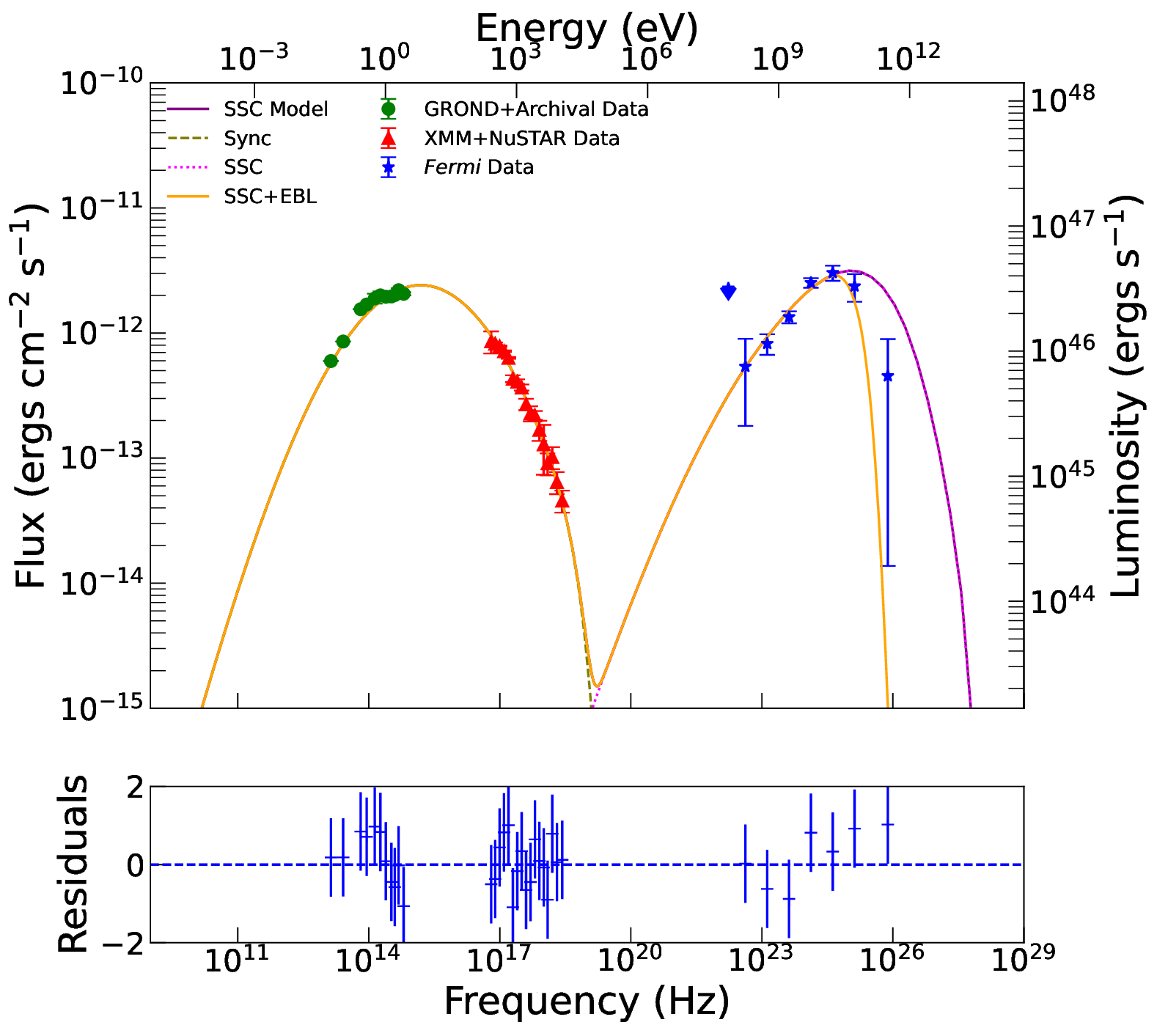}}
		\caption{\label{fig:SSC}Broadband SED of J1520 using GROND, {\it XMM-Newton,  NuSTAR}, and {\it Fermi}-LAT data, modeled using the one-zone leptonic SSC model. The red triangles are the points obtained from XMM+NuSTAR data (binned), while the blue stars are extracted from the fourth {\it Fermi}-LAT catalog and green circles are from GROND and archival data. The total SSC fit is shown by purple solid line. The absorption due to the EBL is taken into account in orange solid line. The synchrotron component is denoted by olive green dashed line and the SSC alone is denoted by the magenta dotted line. The residuals of the fit are also shown, with error $=\pm 1$, for comparison.}
  \label{Fig:SSC}
\end{figure}

The low-energy portion of the SED ($\nu<10^{20}\,\rm Hz$, Figure~\ref{Fig:SSC}) is explained by synchrotron losses of the electron distribution described by Equation~\ref{eq:ele_dis_llp} due to a uniform magnetic field ($B$) encompassing the blob. For the high-energy part ($\nu>10^{20}\,\rm Hz$), we consider two scenarios: (a) the emission can be fully explained by the synchrotron self-Compton losses (SSC, BL Lac-like); (b) the emission requires contribution from external Compton processes (SSC+EC, FSRQ-like). The SSC is mainly driven by parameters such as the magnetic field strength, the bulk Lorentz factor ($\Gamma$) of the electrons, the shape of the particle population, and the size of the emission region ($\rm R$). In the EC scenario, the location of the emission region also plays an important role in determining the observed spectrum. The closer to the black hole, the more the emission would be dominated by EC from the accretion disk photon field, while further along the jet photon fields such as the broad line region clouds and/or the torus region would dominate. 

To minimize possible degeneracies in the fit, several parameters are kept frozen, and/or are linked to one another:
\begin{enumerate}
    \item Black hole mass and accretion disk luminosity: since the non-thermal emission from the jet dominates at IR up to UV frequencies, we do not have any constraints on the black hole mass or disk luminosity of the AGN. The black hole mass is set frozen to standard literature values for objects of a similar nature, $M_{\rm BH}=5\times10^8\,\rm M_{\odot}$ \citep[e.g.][]{Sbarrato_2012, Ghisellini_2012}. The accretion disk luminosity is derived from the empirical relation between the $\gamma$-ray luminosity of the source and the broad line region luminosity \citep{Sbarrato_2012}: $\log(L_{\rm BLR})= 0.93\log(L_{\gamma})+0.63$ and $L_{\rm BLR}= 0.1L_{\rm disk}$. 
    Considering the redshift of our source, we obtain $L_{\gamma}=1.4(\pm0.3)\times10^{47}\,\rm erg~s^{-1}$, which results in $L_{\rm disk}\sim2.84\times10^{45}\,\rm erg~s^{-1}$.
    \item Jet opening angle, distance, and size of the emission region: the jet opening angle ($\theta_{\rm open}$) is 
    fixed at a value of $6^{\circ}$. The size of the emission region ($R$) is linked to the distance from the black
    hole ($R_{\rm diss}$) and $\theta_{\rm open}$ by the following relationship: 
    \begin{equation}
        R = R_{\rm diss}\tan(\theta_{\rm open})
    \end{equation}
    where $R_{\rm diss}$ is a free parameter of the fit.
    \item Size of the BLR and torus regions: the size of the clouds surrounding the black hole are linked to the 
    luminosity of the accretion disk. Following \citet{Ghisellini2009}, we assume that the inner radius of the BLR region, 
    $R_{\rm in, BLR}=10^{17}\sqrt{(L_{\rm disk}/10^{45})}\,\rm cm $, and the outer radius, $R_{\rm out, BLR}=1.1\times R_{\rm in, BLR}$. The size of the torus
    region is related to the accretion disk luminosity by $R_{\rm DT}=2.5\times10^{18}\sqrt{(L_{\rm disk}/10^{45})}\,\rm cm $. 
    \item The accretion disk is modeled as a multi-color black-body emitter; the BLR and torus as a monochromatic black-body emitter, peaking, respectively, at the $H_{\alpha}$ frequency ($\nu_{\rm peak, H\alpha}=4.5\times10^{15}\,\rm Hz$) and  at the typical torus temperature $T_{\rm DT}=900\,\rm K$ ($\nu_{\rm peak, DT}=5.2 \times 10^{13}\,\rm Hz$);
    \item The ratio of cold to electron gas density ($N_{\rm H}$) is kept frozen to the default value of 0.1;    
    \item The accretion efficiency is frozen to 0.1;
    \item Extragalactic background light (EBL) absorption: the EBL (i.e.~all the infrared, optical and UV radiation coming from stars and AGNs since the epoch of reionization, e.g.~\citealp{Hauser2001,Driver2008}) interacts with $\gamma$-ray photons via photon-photon annihilation and pair production. The EBL absorption starts to dominate above $E=10\,\rm GeV$ ($\nu=10^{25}\,\rm Hz$), which 
    is visible as a softening in the Fermi-LAT data. We add the 
    EBL absorption to the fit using the latest model from \citet{dominguez_2023} 
    evaluated at the cosmological redshift $z=1.46$.
    \end{enumerate}
    
The SED fitting is performed using the public code JETSET \citep{Tramacere_2009, Tramacere_2011, Tramacere_2020}, which uses \texttt{Minuit} and Monte-Carlo Markov-Chain (MCMC) optimization method to obtain the best-fit parameters. We add a systematic error of $10\%$ to the multi-wavelength data.



\begin{figure}[ht!]
    \centering
	\makebox{\includegraphics[width=\columnwidth]{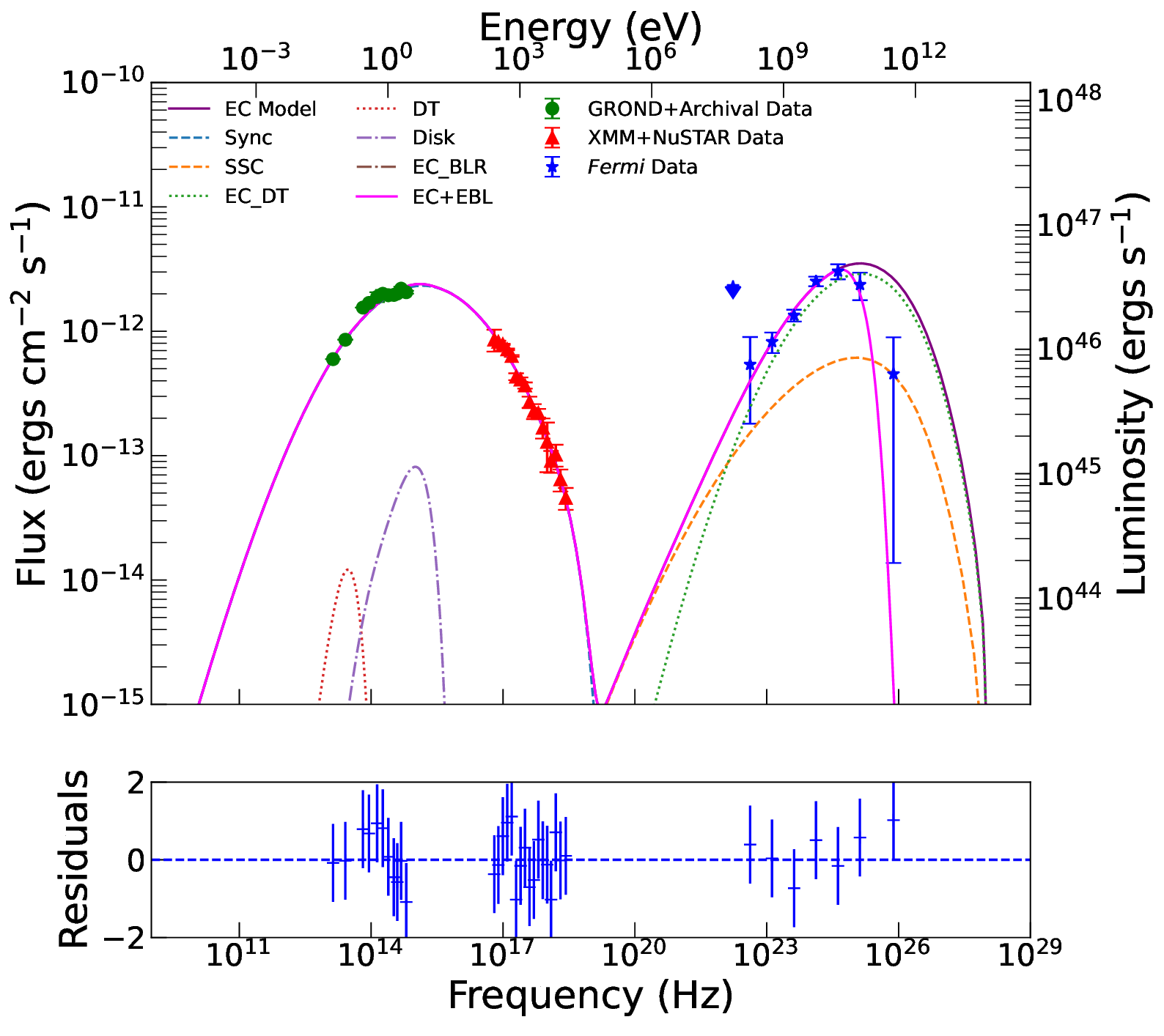}}
		\caption{\label{fig:EC}Broadband SED of J1520 using GROND, {\it XMM-Newton,  NuSTAR}, and {\it Fermi}-LAT data, modeled using EC model. The red triangles are the points obtained from XMM+NuSTAR data (binned), while the blue stars are extracted from the fourth {\it Fermi}-LAT catalog and green circles are from GROND and archival data. Emission region is outside torus region. The contributions of various underlying mechanisms are labelled in plot. Synchrotron component is given in blue dashed line, and SSC alone is given in orange dashed line. The disk is denoted by mauve dash-dotted line and EC due to BLR in brown dash-dotted line. The torus is in red dotted line and its corresponding EC component is in green dotted line. The overall EC fit is denoted by purple solid line. The absorption due to the EBL is taken into account in the pink solid line. The residuals of the fit are also shown, with error $=\pm 1$, for comparison.}
    \label{Fig:EC}
\end{figure}

\begin{figure}[ht!]
    \centering
	\makebox{\includegraphics[width=\columnwidth]{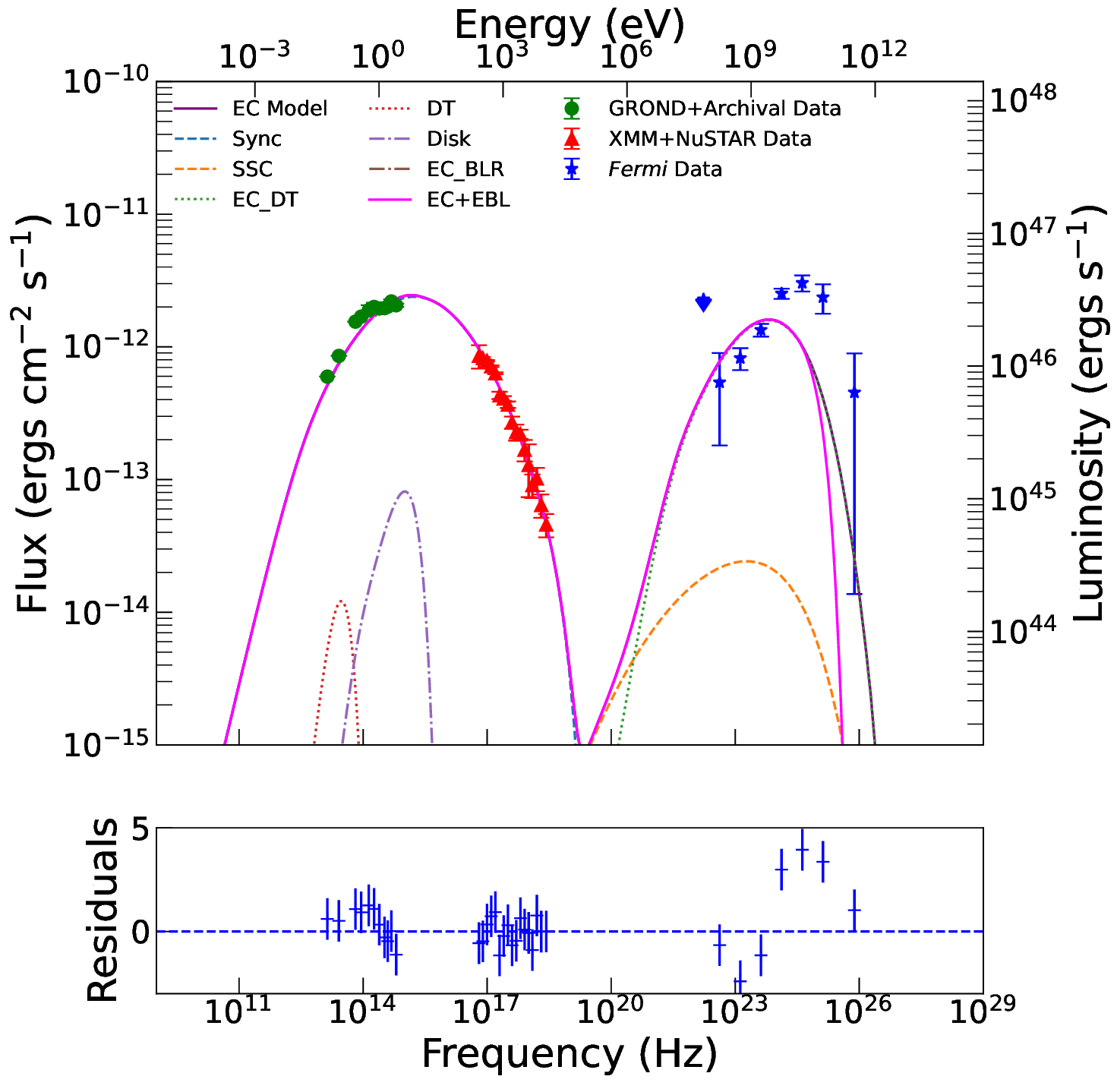}}
		\caption{\label{fig:EC_rdiss_small}Broadband SED of J1520 using GROND, {\it XMM-Newton,  NuSTAR}, and {\it Fermi}-LAT data, modeled using EC model forcing the emission region to the within the torus. The red triangles are the points obtained from XMM+NuSTAR data (binned), while the blue stars are extracted from the fourth {\it Fermi}-LAT catalog and green circles are from GROND and archival data. The contributions of various underlying mechanisms are labelled in plot. Synchrotron component is given in blue dashed line, and SSC alone is given in orange dashed line. The disk is denoted by mauve dash-dotted line and EC due to BLR in brown dash-dotted line. The torus is in red dotted line and its corresponding EC component is in green dotted line. The overall EC fit is denoted by purple solid line. The absorption due to the EBL is taken into account in the pink solid line. The residuals of the fit are also shown, with error $=\pm 1$, for comparison.}
    \label{Fig:pre_EC}
\end{figure}

\subsection{SED fitting results}
The results from both the SSC and the EC models are listed in Table~\ref{Tab:sed_par} and Figures~\ref{Fig:SSC}-\ref{Fig:EC} show the fitted broadband SED. 

The good quality IR/optical to hard X-ray ($E<30\,\rm keV$) data allows to constrain the synchrotron emission on both the SSC only/SSC+EC models. In turn, this is necessary to derive the shape of the particle distribution underlying the emission. Complementary to the optical data, are the $\gamma$-ray data points below $E<10\,\rm GeV$, which are useful to constrain the shape of the low-energy electron distribution in the jet. The derived best-fit spectral shape is consistent in both scenarios, where the slope of the distribution (Eq.~\ref{eq:ele_dis_llp}) is $s_{\rm SSC}=1.09$ and $s_{\rm SSC+EC}=1.61$; the curvature $r_{\rm SSC}=0.71$ 
and $r_{\rm SSC+EC}=0.67$.

The magnetic field values derived in both scenarios are quite low, of the order of $B=10^{-3}\,\rm G$. As for the bulk Lorentz factor, for the SSC scenario, we find it consistent with more normal BL Lacs ($\Gamma_{\rm SSC}=11$), while it is on the high-end for the more canonical FSRQs \citep[$\Gamma_{\rm SSC+EC}=35$, see e.g.][]{Lister_2016, Paliya_2017, Liodakis_2017, Paliya_2019, Lister_2019}. This is in agreement with \citet{Rueda2021_BulkFactor_divide}, who showed that FSRQs, in general, achieve higher bulk Lorentz factors than BL Lacs. The location of the emission region is far along the jet, $R_{\rm diss, SSC}=8.4\times10^{19}\,\rm cm$ ($\sim30 \,\rm pc$) and $R_{\rm diss, SSC+EC}=2.5\times10^{19}\,\rm cm$ ($\sim10 \,\rm pc$). This places the emission region beyond the BLR and the torus region. This means that to reproduce the observed emission, the fit does not need a highly external photon field density. This is mostly relevant in the SSC+EC scenario, as more typical FSRQs require that the emitting region resides within the BLR and/or the torus. We also note that the optical depth of the torus region ($\tau_{\rm DT}=0.1$) is quite low with respect to the standard $\tau_{\rm DT}=0.5$ assumed for FSRQs. This is another indication that the reprocessing of the accretion disk photons required to explain the observed SED is rather low.

As a sanity check, we tested the SSC+EC scenario while forcing the emission region to be within the torus ($R_{\rm diss}<4\times10^{18}\,\rm cm$) and perform the broadband fit. The results are shown in Figure~\ref{fig:EC_rdiss_small}, where it can be seen how this interpretation can be disregarded as the fit falls short of reproducing the high-energy part of the SED. We also tested if a lower black hole mass could be viable for our source, but the results in the fit did not show any improvement.

The total jet power in the two explored models turns out to be consistent with each other, of the order of $P_{\rm tot}=10^{45.5}\,\rm erg~s^{-1}$. In both fitted models, most of the jet power is dominated by the kinetic power of the jet ($P_{k}$), over the radiative throughout of the jet ($P_{r}$). In both the SSC and EC cases, the kinetic power is carried by the electrons ($P_{\rm e}$). We tested that even when we assume one cold proton to one cold electron ($N_{\rm H}=1$) in the jet, the proton power is still subdominant (i.e. one order of magnitude lower) with respect to the electron power. Typically, FSRQs show the opposite behavior, where the kinetic power of the jet is carried by the protons, allowing the jets to extend up to megaparsec scales (akin to typical FRII radio galaxies). Nevertheless, these results are tightly correlated to the shape of the electron distribution, which for J1520 is constrained by the optical and X-ray data. 
The magnetic loading of the jet ($P_{\rm B}$) is also subdominant with respect to the total jet power. Since the best-fit location of the emission region is well beyond the BLR region, the magnetic field decreases significantly, hence the magnetic field density decreases. 

The \citet{dominguez_2023} EBL model fits well the absorption feature at $E>10\,\rm GeV$. In particular, the good quality X-ray data allows us to sample the high-energy electron population shape up to $E\sim20\,\rm keV$, fixing the electron distribution shape without the need of introducing an intrinsic break before the $\gamma_{\rm max}$ of the distribution. Therefore, the EBL attenuation is the major contributor to the high-energy softening of the spectrum.

It is important to note that the source is non-variable in $\gamma$-rays \citep[4FGL-DR4, ][]{4fgl_dr4}. Moreover, on analysing our X-ray data and \textit{Swift/XRT} data, within $0.3-10$ keV (see Section~\ref{sec:target}, \ref{sec:xray}), we obtained consistent flux values, even though the observations were made in different time. This means that the source is also non-variable in X-rays over different epochs. Thus, our SED modeling would not be affected by variability.







\begin{table*}
\begin{center}
\caption{Table of used/derived parameters from the SED of J1520 for the SSC and EC model.}
\hspace{-1.5cm}
\begin{tabular}{lccc}
\hline
Parameter & SSC  &  EC \\
\tableline
\tableline
$reduced-\chi^2$ & 0.63 & 0.62 \\
Black hole mass ($M_{\rm BH}$) in log scale [$M_\odot $]  & --  &  $5\times{10^{8}}$ \\
Accretion disk luminosity ($L_{\rm disk}$) in log scale [erg s$^{-1}$]   & --  & $2.8\times{10^{45}}$ \\
Accretion disk luminosity in Eddington units ($L_{\rm disk}/L_{\rm Edd}$)   & --  & 0.056 \\
Inner radius of the BLR ($R_{\rm BLR}$) [cm] & --  &  $1.68\times{10^{17}}$  \\
Size of the dusty torus ($R_{\rm DT}$) [cm] & --  &  $4.2\times{10^{18}}$  \\
Size of emission region ($R$) [cm] & $8.09\times{10^{18}}$ &  $2.6\times{10^{18}}$ \\
Dissipation distance ($R_{\rm diss}$) [cm] & $8.48(\pm 0.1)\times{10^{19}}$ &  $2.5(\pm0.1)\times{10^{19}}$ \\
Magnetic field ($B$) [G]                                     & $1.0(\pm0.2)\times10^{-3}$ &  $1.03(\pm0.02)\times10^{-3}$ \\
Bulk Lorentz factor ($\Gamma$)                                    & $11.8(\pm0.8)$  & $35.0(\pm9.0)$ \\
Jet viewing angle (in degrees)                                    & $1.85(\pm0.10)$  &   $0.80(\pm0.02)$\\
Low Energy spectral slope ($s$)   & $1.09(\pm0.01)$  & $1.61(\pm0.01)$\\
Spectral Curvature ($r$)   & $0.712(\pm0.005)$ & $0.678(\pm0.006)$\\
Minimum Lorentz factor ($\gamma_{\rm min}$)                       & $1.91(\pm1.34)$  &  $1.17(\pm0.46)$ \\
Turn-over Lorentz factor ($\gamma_{\rm break}$)                       & $1.07(\pm0.02) \times{10^4}$    & $1.29(\pm0.02)\times{10^4}$ \\
Maximum Lorentz factor ($\gamma_{\rm max}$)                       & $9.14(\pm5.04) \times10^6$ &  $5.09(\pm0.09)\times{10^6}$   \\ 
Particle energy density ($U_{e}$) [erg cm$^{-3}$]                & $4.3\times{10^{-6}}$  & $3.7\times{10^{-6}}$\\ 
\hline
\hline
Jet power in electrons ($P_{\rm e}$) in log scale  [erg s$^{-1}$]    & 45.56  & 45.47 \\
Jet power in protons ($P_{\rm p}$) in log scale [erg s$^{-1}$]      & 43.51  & 43.70 \\
Jet power in magnetic field ($P_{\rm B}$) in log scale [erg s$^{-1}$]   & 43.53  & 43.52 \\
Radiative jet power ($P_{\rm r}$) in log scale [erg s$^{-1}$]       & 44.00  &  43.21 \\
Kinetic jet power ($P_{\rm k}$) in log scale [erg s$^{-1}$]       & 45.57 &  45.48 \\
Total jet power ($P_{\rm TOT}$) in log scale [erg s$^{-1}$]         & 45.58  &  45.48  \\
\hline
\end{tabular}
\label{Tab:sed_par}
\end{center}
\end{table*}


\section{Discussion and Conclusion}
\label{sec:conclusion}

The blazar J1520 is indeed a mysterious source. It is a high-$z$ blazar \citep[$z=1.46^{+0.12}_{-0.11}$,][]{kaur2017} that displays both FSRQ and BL Lac type nature. Its featureless optical spectra points towards its BL Lac-like nature \citep{shaw2013b,paiano2017}. It shows a high synchrotron peak frequency $\nu^{sy}_{pk}=1.40(\pm0.08)\times10^{15}$ Hz), similar to BL Lacs. It has a high radio luminosity ($L_{1.4GHz}=8.9(\pm0.2)\times 10^{42}\,\rm erg~s^{-1}$) which is characteristic of FSRQs \citep{Giommi2012a, Giommi2012b}. It possesses high synchrotron peak luminosity ($L_{sy}^{pk}=3.4(\pm0.2) \times10^{46}\,\rm erg~s^{-1}$), like FSRQs. This challenges the `blazar sequence' which states that BL Lacs as luminous as FSRQs, but with synchrotron peak frequencies $\nu^{sy}_{pk}>10^{15}$ Hz, should not exist \citep[eg.][]{BlazaeSeqFossati1998, GhiselliniBlazSeq2008, GhiselliniBlazSeqNew2017}. BL Lacs accelerate particles to very high energies, compared to FSRQs, and hence display high synchrotron peak frequencies \citep[eg.][]{Costamante2001, Tavecchio2011}. On the other hand, in FSRQs particles in the jet upscatters photons from surrounding radiation fields to emit high luminosities \citep{Ghisellini2009}. Since we find both these characteristics in our source, the general mechanism of jet acceleration and emission is brought into question. A possible scenario to explain characteristics of both classes of blazars can be a FSRQs having its optical emission lines swamped by powerful jet. Thus, it is imperative to probe its true nature and its underlying physics. Moreover, due to its significant emission $>$10 GeV, it could be used to test EBL models \citep{Ackermann2012, Dwek2013, Biasuzzi2019}.

On gathering the multiwavelength data of J1520, we are able to construct its SED. Firstly, we use X-ray data to gain knowledge about the underlying distribution of the high-energy electron population. We find that log-parabolic distribution is favored over other models, along with a soft X-ray spectral index, $\Gamma_X$=2.73${\pm0.07}$. Using this information, we venture to model the broadband SED of J1520. We are able to constrain the synchrotron peak using the optical-IR and X-ray data. Gamma ray data gives the shape of high energy peak.

On modeling the SED on J1520 using SSC and EC scenarios, we find both produce good fits. The associated parameters of fit are displayed in Table \ref{Tab:sed_par}. The SSC model gives a good fit to the data as shown in Fig. \ref{Fig:SSC}. The EC model gives a good fit, as shown in Fig. \ref{Fig:EC}, and reconciles with the observation of high radio power and high synchrotron peak luminosity. In this case, the emission region ($R_{diss}$) is located outside the BLR and torus region. This is in agreement with previous studies on `blue FSRQs' by \citet{blueFSRQ} and \citet{Padovani2012}. The electrons are not cooled efficiently by external photon fields, since the emission region is at $\sim$10pc, well beyond the BLR and torus. In this way, they can get accelerated to high energies, giving a high synchrotron peak frequency. The radiation field from torus region interacts with the jet particles, but are not Doppler boosted, thus it can emit luminosities higher than typical of BL Lacs. For comparison, we also fit an EC model, forcing the emission region to be inside the torus region. Figure \ref{Fig:pre_EC} shows the result of the fit. We find that such a constraint cannot explain the data.

\begin{figure}[tb!]
    \centering
    \makebox{\includegraphics[width=\columnwidth]{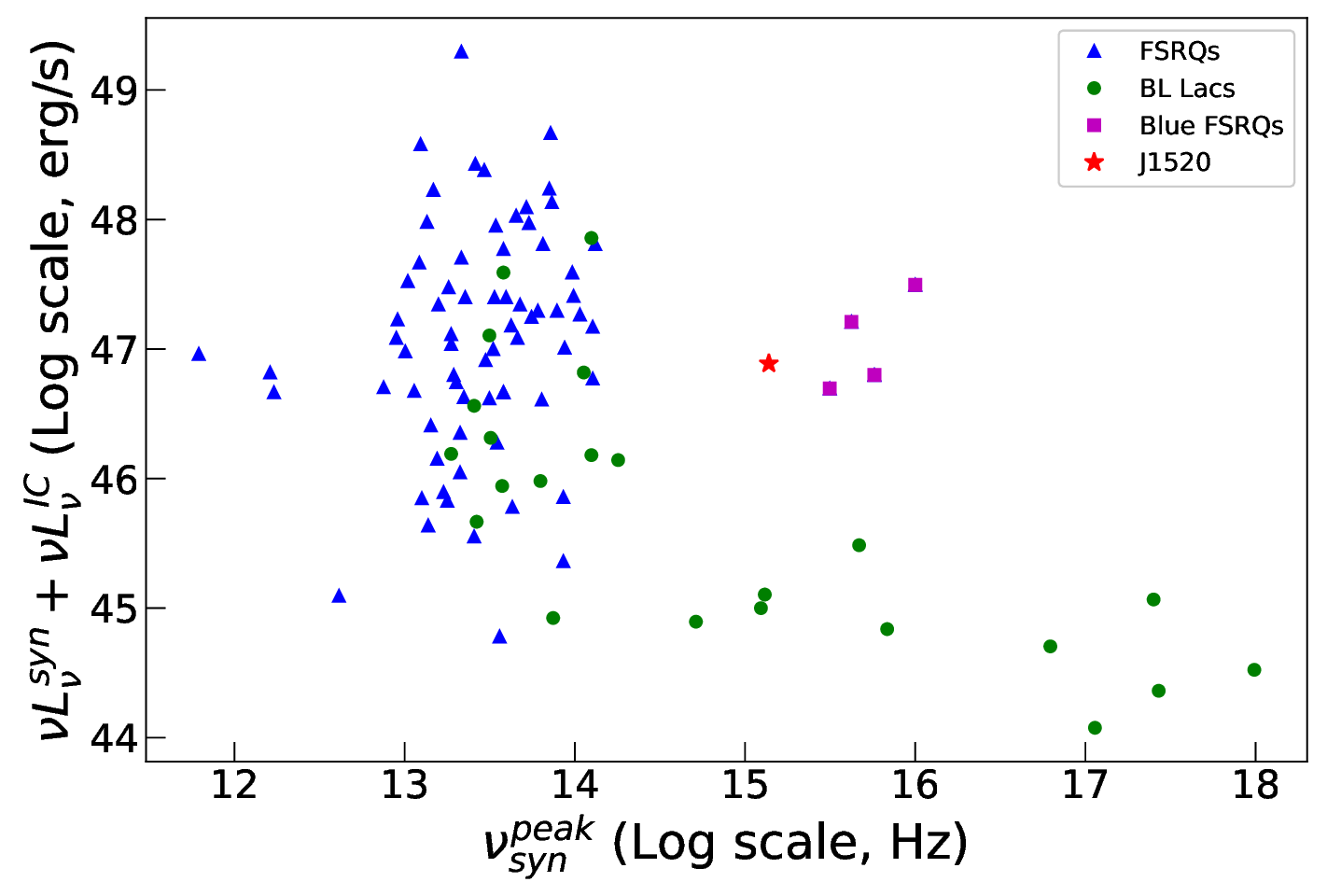}}	
		\caption{\label{fig:padovani2012} The plot shows typical positions of FSRQs and BL Lacs in a $\nu_{syn}^{peak}-L^{peak}$ plane. Blue triangles denote FSRQs, green circles are BL Lacs (both datasets come from \citet{Padovani2012}). Pink squares are previously found `blue FSRQs' \citep[]{blueFSRQ,Padovani2012} and the red star is our source J1520.
		}
\end{figure}

Figure \ref{fig:padovani2012} shows blazars (FSRQs as blue triangles and BL Lacs as green circles), including our source, on $\nu_{syn}^{peak}-L^{peak}$ plane ($L^{peak}$ is represented by the sum of the synchrotron and inverse-Compton peak luminosities). 
We see that our source J1520 (indicated by red star) is an outlier with respect to FSRQs and BL Lacs. It lies near 4 `blue FSRQs' (pink squares), given in \citet{blueFSRQ}. This strengthens the case that J1520 is a `blue FSRQ' or `masquerading BL Lac', whose optical features have been overpowered by jet emission. The classification scheme proposed by \citet{Ghisellini2011}, separates radiatively efficient disks in FSRQs from inefficient ones in BL Lacs according to the $L_{disk}/L_{Edd}$ ratio. If $L_{disk}/L_{Edd}>0.005$, it falls in the radiatively efficient regime of FSRQs. Moreover, \citet{TXS0506+056} suggests that blue FSRQs have $L_{disk}/L_{Edd}>0.01$. The $L_{disk}/L_{Edd}=0.056$ (Table \ref{Tab:sed_par}) of J1520 hints towards it being in the blue FSRQ regime. Furthermore, \citet{Sbarrato2012} proposed a similar divide using $L_{\gamma}/L_{Edd}>0.1$ for FSRQs. J1520 shows $L_{\gamma}/L_{Edd}=3.1$ pointing towards its FSRQ-like nature.

To explain the cutoff in the high energy $\gamma$-ray data, we use the EBL model given in \citet{Saldana-Lopez2021} and \citet{dominguez_2023, Dominguez2023b}. The EBL model explains fairly well the emission that we observe at E$>$10 GeV. High power high-z sources like J1520 can be useful probes for constraining EBL models since they exhibit a curvature in the data due to interaction with EBL. Since redshift measurements for BL Lacs type sources, with flat optical continuum, is challenging, EBL absorption can pose an effective method to estimate their redshifts \citep{dominguez_2023}.

The evidence provided in this study highlight J1520 as a new ``blue FSRQ", increasing the population of this mysterious source class to seventeen. Studying more candidates of such sources with X-ray instruments such as XMM, Swift and NuSTAR is paramount to understand the physics behind these objects.

\vspace{5cm}
We thank the referee for comments that helped in improving the manuscript. This research has made use of data from the $NuSTAR$ mission, a project led by the California Institute of Technology, managed by the Jet Propulsion Laboratory, and funded by the National Aeronautics and Space Administration (NASA). Data analysis was performed using the $NuSTAR$ Data Analysis Software (NuSTARDAS), jointly developed by the ASI Science Data Center (SSDC, Italy) and the California Institute of Technology (USA).

This research has made use of data based on observations obtained with $XMM-Newton$, an ESA science mission with instruments and contributions directly funded by ESA Member States and NASA.

Part of this work is based on archival data, software and online services provided by the Space Science Data Center - Italian Space Agency (ASI).

This research has made use of data and software provided by the High Energy Astrophysics Science Archive Research Center (HEASARC), which is a service of the Astrophysics Science Division at NASA/Goddard Space Flight Center (GSFC).

GR and MA acknowledge funding under NASA contract 80NSSC22K1807. 

Support for this work was provided by NASA through the NASA Hubble Fellowship grant \#HST-HF2-51486.001-A 
awarded by the Space Telescope Science Institute, which is operated by the Association of Universities 
for Research in Astronomy, Inc., for NASA, under contract NAS5-26555.






\end{document}